\documentclass[12pt,preprint]{aastex}

\usepackage{color}

\begin{document}

\title{A statistical model for the $\gamma$-ray variability of the Crab nebula}

\author{Qiang Yuan$^{1,2}$, Peng-Fei Yin$^1$, Xue-Feng Wu$^{2,3,4}$,
Xiao-Jun Bi$^{1}$, Siming Liu$^3$ and Bing Zhang$^2$}

\affil{
$^1$Key Laboratory of Particle Astrophysics, Institute of High Energy
Physics, Chinese Academy of Sciences, Beijing 100049, P. R. China\\
$^2$Department of Physics and Astronomy, University of Nevada Las Vegas,
Las Vegas, NV 89154, USA\\
$^3$Purple Mountain Observatory, Chinese Academy of Sciences, Nanjing
210008, P. R. China\\
$^4$Joint Center for Particle Nuclear Physics \& Cosmology (J-CPNPC),
Nanjing 210093, P. R. China
}

\begin{abstract}

A statistical scenario is proposed to explain the $\gamma$-ray
variability and flares of the Crab nebula, which were observed
recently by the Fermi/LAT. In this scenario electrons are
accelerated in a series of knots, whose sizes follow a power-law
distribution. These knots presumably move outwards from the pulsar and
have a distribution in the Doppler boost factor. The maximal electron
energy is assumed to be proportional to the size of the knot. Fluctuations
at the highest energy end of the overall electron distribution will result
in variable $\gamma$-ray emission via the synchrotron process in
the $\sim 100$ MeV range. Since highly boosted larger knots are rarer
than smaller knots, the model predicts that the variability of the
synchrotron emission increases with the photon energy. We realize such
a scenario with a Monte-Carlo simulation and find that the model can
reproduce both the two $\gamma$-ray flares over a period of
$\sim$ year and the monthly scale $\gamma$-ray flux fluctuations
as observed by the Fermi/LAT. The observed $\gamma$-ray spectra
in both the steady and flaring states are also well reproduced.

\end{abstract}

\keywords{radiation mechanism: non-thermal --- pulsar wind nebula:
individual: Crab --- gamma rays: theory}

\section{Introduction}

The Crab pulsar wind nebula is powered by its central pulsar born
from a supernova explosion in 1054. It is a very luminous source
in almost all wavelengths, from radio to the very high energy (VHE)
$\gamma$-ray bands. The broadband non-thermal emission spectrum is
well modeled by a
synchrotron-inverse Compton (IC) scenario \citep{1996MNRAS.278..525A,
2010A&A...523A...2M} with the transition from the synchrotron to IC
component occurring at a few hundred MeV \citep{2010ApJ...708.1254A}.
The overall emission seems to be steady, so that the Crab nebula has
been adopted as a standard candle in high energy astrophysics to
calibrate observations from different instruments.

However, detailed images of the Crab nebula in optical and X-ray bands
indicated dynamical structures at small scales. Observations
by the Hubble Space Telescope revealed wisps and knots in the nebula,
and resolved some substructures of the highly variable wisps
\citep{1995ApJ...448..240H}. X-ray observations by ROSAT and Chandra
uncovered a jet-torus structure of the inner nebula
\citep{1995ApJ...448..240H,2000ApJ...536L..81W}, and the equatorial
ring is found moving outwards with a speed of $\sim 0.5c$
\citep{2002ApJ...577L..49H}. At higher energies in the $\gamma$-ray band,
detailed structures of the nebula generally can not be resolved. However,
COMPTEL and EGRET observations indicated that
the synchrotron radiation in $1-150$ MeV is variable on a timescale of
$\sim1$ yr \citep{1995A&A...299..435M,1996ApJ...457..253D}.

In September 2010, AGILE collaboration reported a $\gamma$-ray
flare above $100$ MeV from the direction of the Crab nebula, which
lasted for about 3 days (ATel \#2855; Tavani et al. 2011). The flux 
during the flare period is about $2-3$ times higher than the average one. 
This flare was soon confirmed by the Fermi/LAT collaboration (ATel
\#2861), and archive search of the Fermi/LAT data revealed another
flare in February 2009 \citep{2010arXiv1011.3855F}. There were many
other simultaneous or follow-up measurements for the flare in
September 2010 in X-ray, optical, infrared and radio bands.
However, no significant flux enhancement was discovered. In the
VHE $\gamma$-ray energies, ARGO-YBJ collaboration claimed the
detection of a flux enhancement around TeV, with a possibly longer
duration (ATel \#2921). However, MAGIC and VERITAS observed the
source with a shorter duration during the flare phase and did not
reveal any enhancement in flux (ATel \#2967, \#2968).

It was proposed that the flares was due to synchrotron emission of
ultra-relativistic electrons with energies up to $\sim$PeV
\citep{2010arXiv1011.3855F}. Considering the fact that 
the rest frame synchrotron radiation in a magnetic field dominated 
acceleration regime can not exceed $\sim 70$ MeV due to the fast 
synchrotron cooling of high energy electrons \citep{1970RvMP...42..237B,
2010MNRAS.405.1809L}, the observed GeV emission implies Doppler shift 
of the radiation region 
or other acceleration mechanisms instead of shock acceleration
\citep{2010arXiv1011.3855F}. \cite{2010arXiv1011.1800K} proposed
that the $\gamma$-ray variability (flare) originated from the
``inner knot'' of the Crab nebula (``knot 1'' as defined in
\citet{1995ApJ...448..240H}), with mildly Doppler-boosted emission.
The instability of the termination shock may cause the
variability as revealed by magnetohydrodynamic (MHD) simulations.
\cite{2010arXiv1011.4176B} suggested that electrons are accelerated in a
region behind the shock, and the variability was attributed to changes
in the maximum energy of accelerated electrons, electron
spectral index or the magnetic field.

In this Letter we employ a statistical approach to model the
$\gamma$-ray variability of the nebula. Fermi/LAT observations have shown
that the low-energy synchrotron component is variable on monthly
time scale, while the high-energy IC component seems to be stable
\citep{2010arXiv1011.3855F}.
Since the synchrotron $\gamma$-rays are produced by the highest energy
electrons, these observations indicate that fluctuations at the
high-energy end of the electron distribution might be responsible for the
variability. It is natural to expect that events that can generate
the highest energy electrons are rarer, and therefore would lead to
the largest fluctuation. Therefore the variability and flares in the sub-GeV
$\gamma$-rays can be simply due to the statistical fluctuation of the
highest energy electrons achievable in the electron accelerators.
Lower energy electrons do not suffer from significant fluctuations
since many more accelerators can contribute to them simultaneously,
which gives rise to a ``steady-state'' emission in both the lower energy
synchrotron component and the higher energy IC component.

\section{The Model and Results}

We build a physically possible model to investigate the above
picture in more details. We assume that electrons are
accelerated in a series of knots with a power-law energy spectrum
$F_i(E) \propto E^{-\alpha_e}$ with the maximum electron energy
$E_{\rm max}^i$ being proportional to the size of the $i$th 
knot\footnote{The maximum energy of electrons will reach a critical
value when the synchrotron cooling is essential. The critical Larmor
radius is estimated as $R_L^c\approx \sqrt{\frac{6\pi e}{\sigma_TB}}
\frac{m_ec^2}{eB}\simeq 2\times 10^{-3}(B/{\rm mG})^{-1.5}$pc, which 
just corresponds to the size of the emission region as indicated by the 
day-scale flare. Therefore we assume the largest size
of the knots is $R_L^c$, and for all other knots smaller than $R_L^c$ 
the scaling relation between $E_{\rm max}^i$ and $r_i$ can well hold.}. 
We further assume that the size of knots has a
power-law distribution $P(r_i)\propto r_i^{-\beta}$. Adding the
contribution from all these knots together, noticing that the
normalization of the synchrotron spectrum of the knots scales as
$r_i^3$, we get the total electron spectrum as
\begin{eqnarray}
F(E)&\propto&\int F_i(E)\Theta(E_{\rm max}^i-E)\times r_i^3P(r_i)
{\rm d}r_i\nonumber\\
&\propto& \int E^{-\alpha_e}\Theta(E_{\rm max}^i-E) \times 
{E_{\rm max}^i}^{3-\beta}{\rm d}E_{\rm max}^i\nonumber\\
&=&E^{-\alpha_e}\int_E^{E_{\rm max}}{E_{\rm max}^i}^{3-\beta}{\rm d}
E_{\rm max}^i,
\end{eqnarray}
where $\Theta(x)$ is the Heaviside step function, $E_{\rm max}$ is
the maximum energy of the largest knot. If $\beta>4$, the above 
integral of the last step is approximately proportinal to 
$E^{4-\beta}$ given the maximum energy of the largest knot 
$E_{\rm max}\gg E$; if $\beta<4$ the integral is nearly a constant
determined by $E_{\rm max}$. Therefore the electron spectrum is
a power law $F(E)\propto E^{-\alpha}$ with power-law index
$\alpha=\max (\alpha_e+\beta-4,\alpha_e)$ and a cutoff determined 
by the largest knot.

The maximal energy of synchrotron emission in the magnetic field dominated
acceleration regime is $\sim 70$ MeV \citep{1970RvMP...42..237B} (defined 
by equating the synchrotron cooling time scale to the gyro-period,
the minimum of the acceleration time scale). For smaller accelerators,
the maximum energy is also limited by the size of the accelerator.
In order to account for the observations in the $\sim 100$ MeV
range, we therefore further employ a mild Doppler boost factor. We
assume a Gaussian distributed Lorentz factor of all the knots with
a mean value $\Gamma=2.0$ and a standard deviation $\sigma=0.25$.
Such a Lorentz factor is consistent with the upper limit of the
typical velocity of the jet \citep{2010arXiv1011.3855F}. We assume
that the angle $\theta$ between the knot motion and the line-of-sight
is randomly distributed. The Doppler factor
$\delta=1/\Gamma(1-v\cos\theta)$ (where $v$ is the velocity of the
knot in unit of $c$) therefore is distributed within the range
$0.18-5.5$ for a $3\sigma$ distribution of $\Gamma$. Only knots
with large enough size and large enough Doppler factor can
contribute to high energy synchrotron radiation to account for the
flares observed by the Fermi/LAT.

We realize such a picture by a Monte-Carlo simulation. We directly
simulate the synchrotron spectrum in each knot instead of the
electron spectrum. The synchrotron spectrum of each knot in its
co-moving frame is adopted as $\nu'
F_{\nu'}\propto\nu'^{-\alpha_{\nu}} \exp(-\nu'/\nu'_{\rm max})$,
where $\nu'_{\rm max} \propto(E_{\rm max}^i)^2 \propto r_i^2$,
$\alpha_{\nu}=(\alpha-3)/2$ is the synchrotron spectral index.
where $\alpha$ is the power-law index of the overall electron distribution.
The magnetic field in each knot is not explicitly
used in the calculation. The magnetic field will affect the cooling
break of the low energy synchrotron spectrum as well as the high energy
cooling time scale. To keep the basic framework of the present study, 
we may expect the magnetic field not to vary significantly with respect 
to the knot size.

\begin{figure}[!htb]
\centering
\includegraphics[width=0.7\columnwidth]{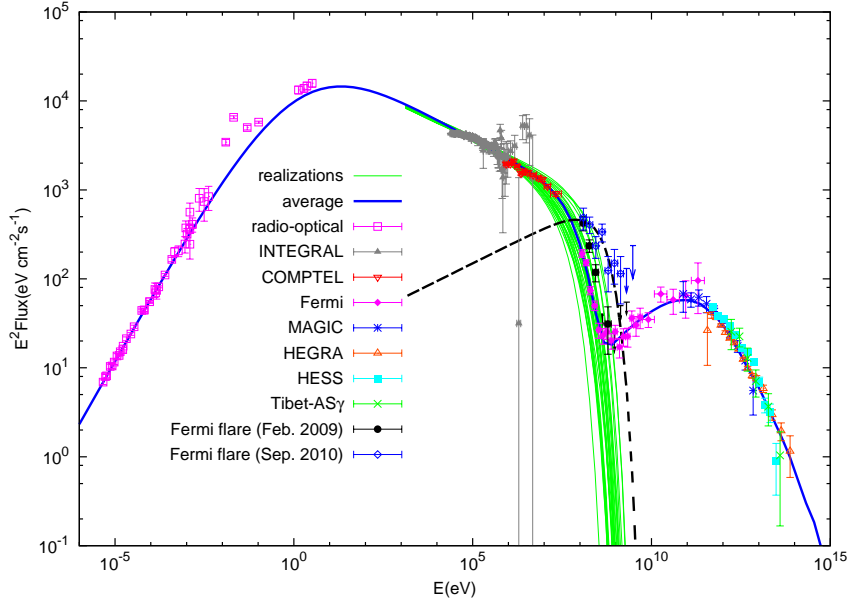}
\caption{Spectral energy distribution of the Crab nebula. The thin green
lines denote the simulated synchrotron spectra of $30$ realizations. The
thick blue line denote the fit to the multi-wavelength steady state
observational data, which can be understood as the average result of many
realizations in our model. The black-dashed curve denotes the contribution 
from a maximum size knot with slightly specific parameters (see the text), 
which is shown to be able to explain the large flare in September 2010. 
The references
of the observational data are: radio-optical \citep{2010ApJ...711..417M},
INTEGRAL \citep{2009ApJ...704...17J}, COMPTEL \citep{2001A&A...378..918K},
Fermi/LAT \citep{2010ApJ...708.1254A}, MAGIC \citep{2008ApJ...674.1037A},
HEGRA \citep{2004ApJ...614..897A}, HESS \citep{2006A&A...457..899A},
Tibet-AS$\gamma$ \citep{2009ApJ...692...61A}, and Fermi/LAT flares
\citep{2010arXiv1011.3855F}.
}
\label{fig:f1}
\end{figure}

The energy loss of electrons in the knots can be important and needs to be
considered. For instantaneous injection of accelerated electrons, the
probability of observing electrons at a given energy is proportional to
the corresponding synchrotron cooling time $t_c\propto E^{-1}$. For low
energy electrons with cooling time longer than the age of the Crab
nebula $t_{\rm age} \approx 10^3$ yr, these electrons will survive and
accumulate. Therefore we can construct a probability function
$P(E)=1/(1+t_{\rm age}/t_c)$, describing the effect of cooling on
the electron spectrum. The cooling time of electrons producing
synchrotron photon with energy $\epsilon$ is $t_c\simeq 1.5(B/{\rm
mG})^{-1.5}(\epsilon/{\rm keV})^{-0.5}\delta^{-0.5}$ yr. For an average
magnetic field $B\approx 0.1$ mG, a mildly Doppler factor $\delta\sim 1$, 
electrons producing photons with an energy $\epsilon\sim 2.5$ eV have a 
cooling time comparable to $t_{\rm age}$. The observed spectrum of Crab 
nebula has a break near $2.5$ eV (see below Fig. \ref{fig:f1}).

In this work we mainly focus on the high energy part of the
spectra, e.g., from X-ray to $\gamma$-ray band, then $P(E)\propto
E^{-1}$. Assuming that the accelerated electrons have a spectrum index
$\alpha_e^{\rm inj}$, the electron spectrum index in the synchrotron
cooling dominated regime is then $\alpha_e=\alpha_e^{\rm inj}+1$.
Here we adopt $\alpha_e=2.6$ which corresponds to $\alpha_e^{\rm inj} =
1.6$, as required by fitting the radio-optical data 
\citep{2010A&A...523A...2M}. The maximum value of $\nu'_{\rm max}$ is 
taken as $55$ MeV in the co-moving frame (corresponding to the largest 
knot). This value is derived according to a fit to the 
average Fermi/LAT spectrum of the synchrotron component 
\citep{2010ApJ...708.1254A}, given the Lorentz factor distribution. 
The power-law index of the size distribution is adopted as $\beta\approx4.8$.
Adding all the contribution from the knots together we can get a
total synchrotron spectrum $\nu'F_{\nu'}\propto\nu'^{-0.2}$, which
can reproduce the observed optical-MeV band data of the Crab
nebula (Fig. \ref{fig:f1}). Note that these parameters, 
i.e., $\Gamma$, $\nu'_{\rm max}$, $\alpha_e$ and $\beta$ are not uniquely
determined. What we adopt in this work is an illustration of the model. 
Once the data from optical to $\gamma$-ray band are reproduced, the 
discussion below is basically unchanged if a different set of parameters
is taken.

For each knot we randomly generate an angle $\theta$ and a Lorentz
factor $\Gamma$ based on the assumed Gaussian distribution, and
calculate the Doppler factor $\delta$. Then the frequency of 
synchrotron emission from each knot is shifted from $\nu'$ to
$\delta\nu'$, and the flux is enhanced to $\delta^3F_{\nu'}$. The
contribution from all the knots are summed up to obtain
the total emission spectrum.

A series of simulated synchrotron spectra are shown in Fig. \ref{fig:f1},
which are compared with the observational data of the Crab nebula. Note
that for the INTEGRAL data we adopt a flux normalization factor $0.6$ in
order to be consistent with the extrapolation of COMPTEL data. In
\cite{2010A&A...523A...2M} a $0.78$ factor was adopted to match the
XMM-Newton data. The difference of flux normalization may be due to
discrepancies in calibration of different detectors. The fact that
the INTEGRAL data may contain emission from the Crab pulsar may also
contribute to the difference. For comparison we show the fitting results
to the broad band data with the blue thick line, with a broken
power-law electron spectrum. This result is similar to the model invoking
two population electrons as introduced in \cite{1996MNRAS.278..525A} and
\cite{2010A&A...523A...2M}. It is shown that the simulated synchrotron 
spectra have very small fluctuations in the X-ray energy band, but fluctuate
significantly at higher energies ($>1$ MeV). The average of the simulated
$\gamma$-ray flux would be consistent with the long term Fermi/LAT and
COMPTEL data. The fluctuations can be responsible for the observed
variabilities and flares. 

The very large flare at September 2010 that is not well 
reproduced by these simulations can be regarded as an event with very 
small probability. Such an event may be due to a large knot with specific 
parameters different from what used in the model. The black-dashed line in 
Fig. \ref{fig:f1} shows a possible reproduction of the large flare, 
produced by a knot with maximum size, Doppler factor $\delta=5.5$, cutoff 
energy of synchrotron radiation in its comoving system $70$ MeV and
an additional normalization factor $0.2$ of the flux. Detailed modeling 
of the knot formation and its evolution is expected to better address
this issue.

\begin{figure}[!htb]
\centering
\includegraphics[width=0.8\columnwidth]{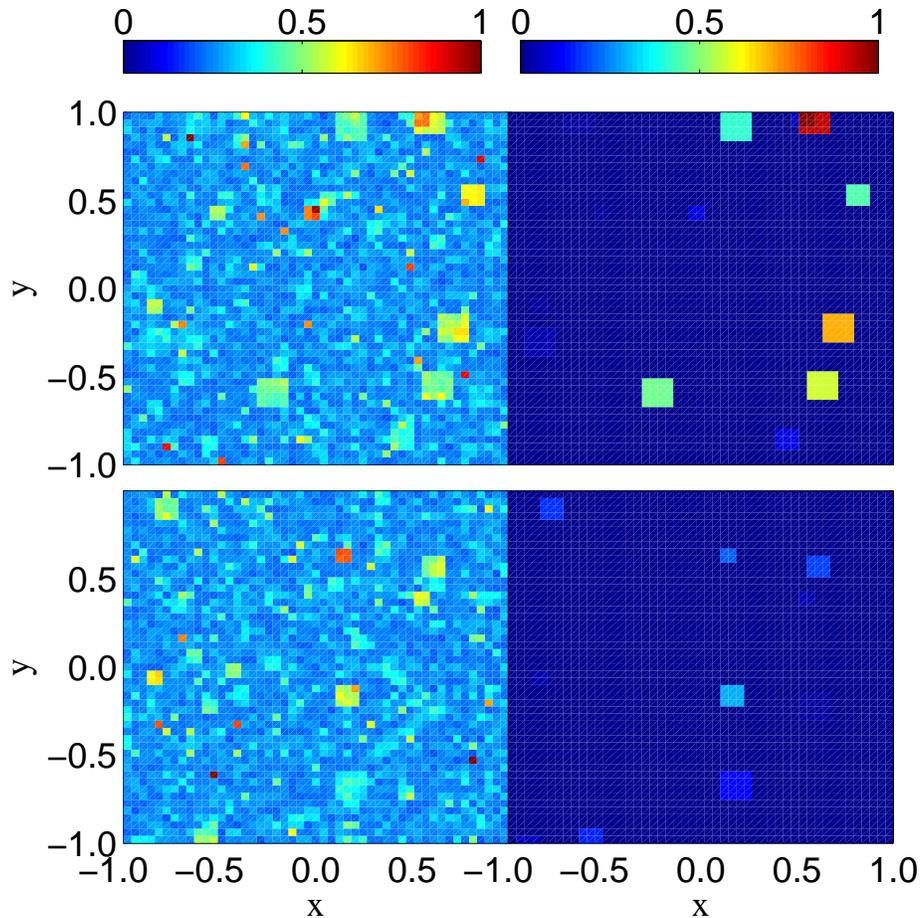}
\caption{Upper and bottom panels represent two realizations of the
synchrotron sky-maps at 10 keV (left) and 100 MeV (right) energies.
The upper one shows a realization of a possible $\gamma$-ray flare,
while the lower one shows a realization with relatively weak 
$\gamma$-ray emission.
}
\label{fig:f2}
\end{figure}

We pick out two realizations and show the synchrotron sky-maps at
10 keV (left) and 100 MeV (right) in Fig. \ref{fig:f2},
respectively. To generate a sky-map, we designate an
$x-y$ plane with $50 \times 50$ pixels. We randomly assign a
position for each knot in the $x-y$ plane, with the largest knot
having a size of $5\times 5$ pixels. We then sum up contributions 
of all the knots located in each pixel and obtain the sky-map. 
It is shown that in the X-ray band (left panels) the 
skymaps are relatively smooth with some hot spots due to several large 
knots. These hot spots may correspond to the Chandra and Hubble knots of
the Crab nebula (ATel \#2882, \#2903). The total flux of X-ray emission,
however, is statistically steady from time to time.
The width of the total X-ray flux distribution at $\sim 10$ keV is 
estimated to be of the order $1\%$, which is consistent with the upper
limit of $5.5\%$ obtained by Swift/BAT (ATel \#2858). 
The $\sim 100$ MeV $\gamma$-ray sky-map could differ significantly 
from each other. In some cases, one or several large knots would 
dominate the total flux. The predicted energy-dependent spatial scales 
due to fluctuations may not be ready to test with the available high 
energy telescopes, but higher angular resolution observations in the 
future will help.

According to our model, the IC component should not show
significant variability except in the high energy regime ($>100$
TeV), depending on the maximal electron energies. For example, for
a magnetic field $B\approx1$ mG, the maximal Lorentz factor of electron 
can be as high as $\gamma_{\rm max}\sim \sqrt{6\pi e/\sigma_TB}\simeq
4\times 10^9$. Therefore only the IC photons with energies higher than
several hundred TeV would have relatively large fluctuations. This
conclusion is different from that of \cite{2010arXiv1011.4176B},
who suggested that the IC $\gamma$-ray spectra also vary
significantly at multi-TeV energies.

\begin{figure}[!htb]
\centering
\includegraphics[width=\columnwidth]{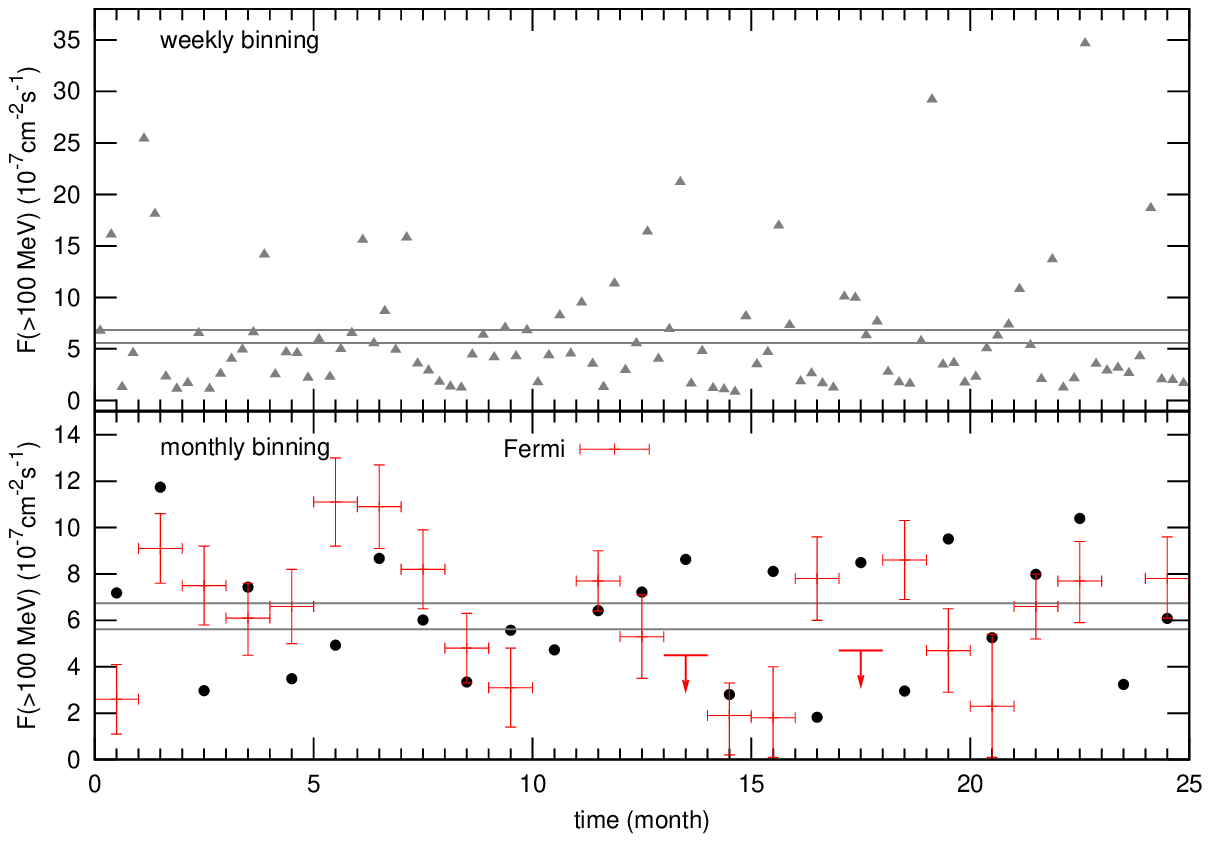}
\caption{Light curves of the simulated synchrotron flux above $100$
MeV. The upper panel shows the results with weekly bin and the
bottom panel for monthly bin. The vertical axes in each panel
indicate the average value of the simulations, which is normalized
to the observational value $\sim 6.2\times 10^{-7}$ cm$^{-2}$
s$^{-1}$ \citep{2010arXiv1011.3855F}. The Fermi/LAT observational
data with monthly bin are also plotted in the lower panel.}
\label{fig:f3}
\end{figure}

\begin{figure}[!htb]
\centering
\includegraphics[width=0.45\columnwidth]{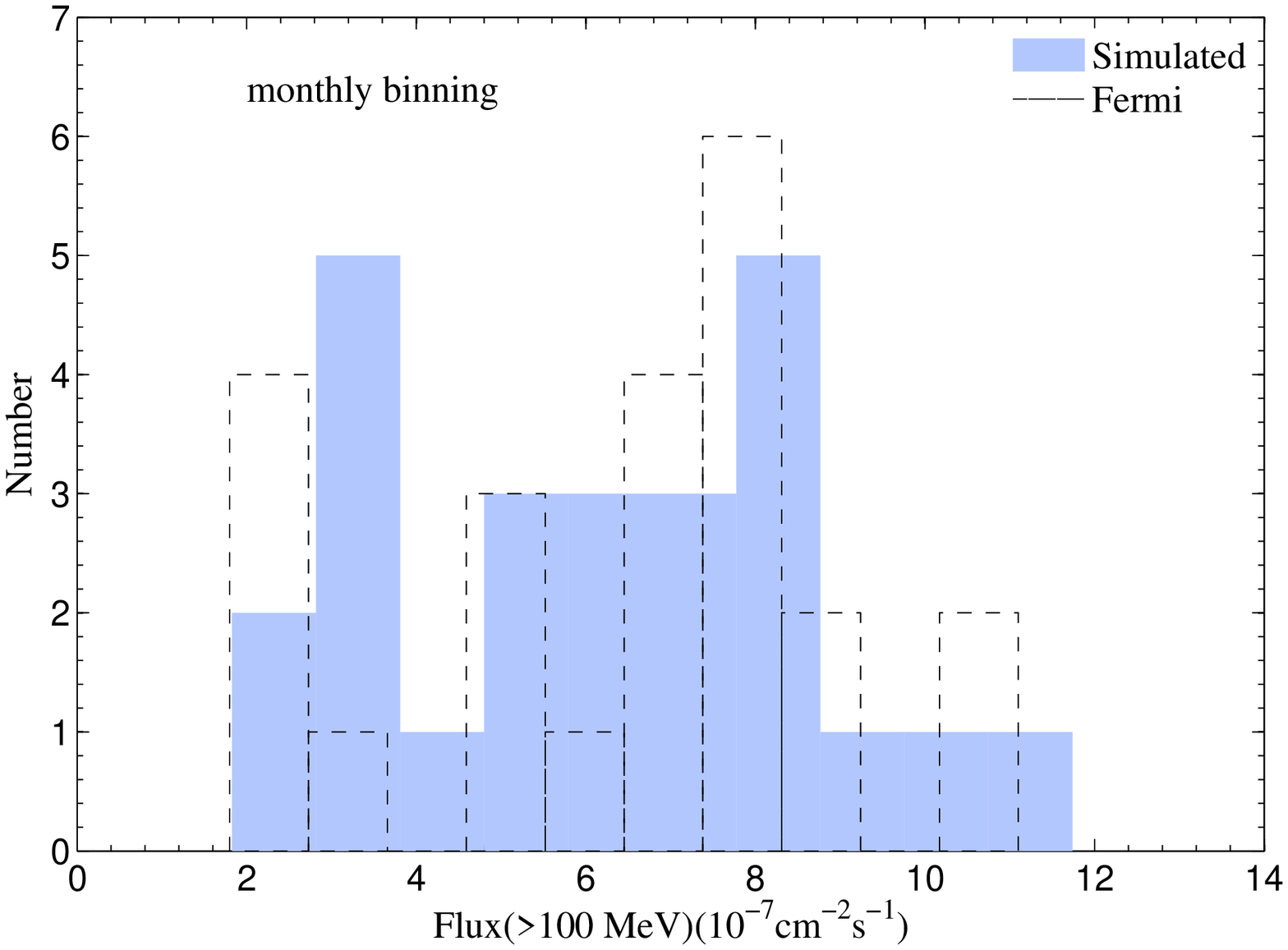}
\includegraphics[width=0.45\columnwidth]{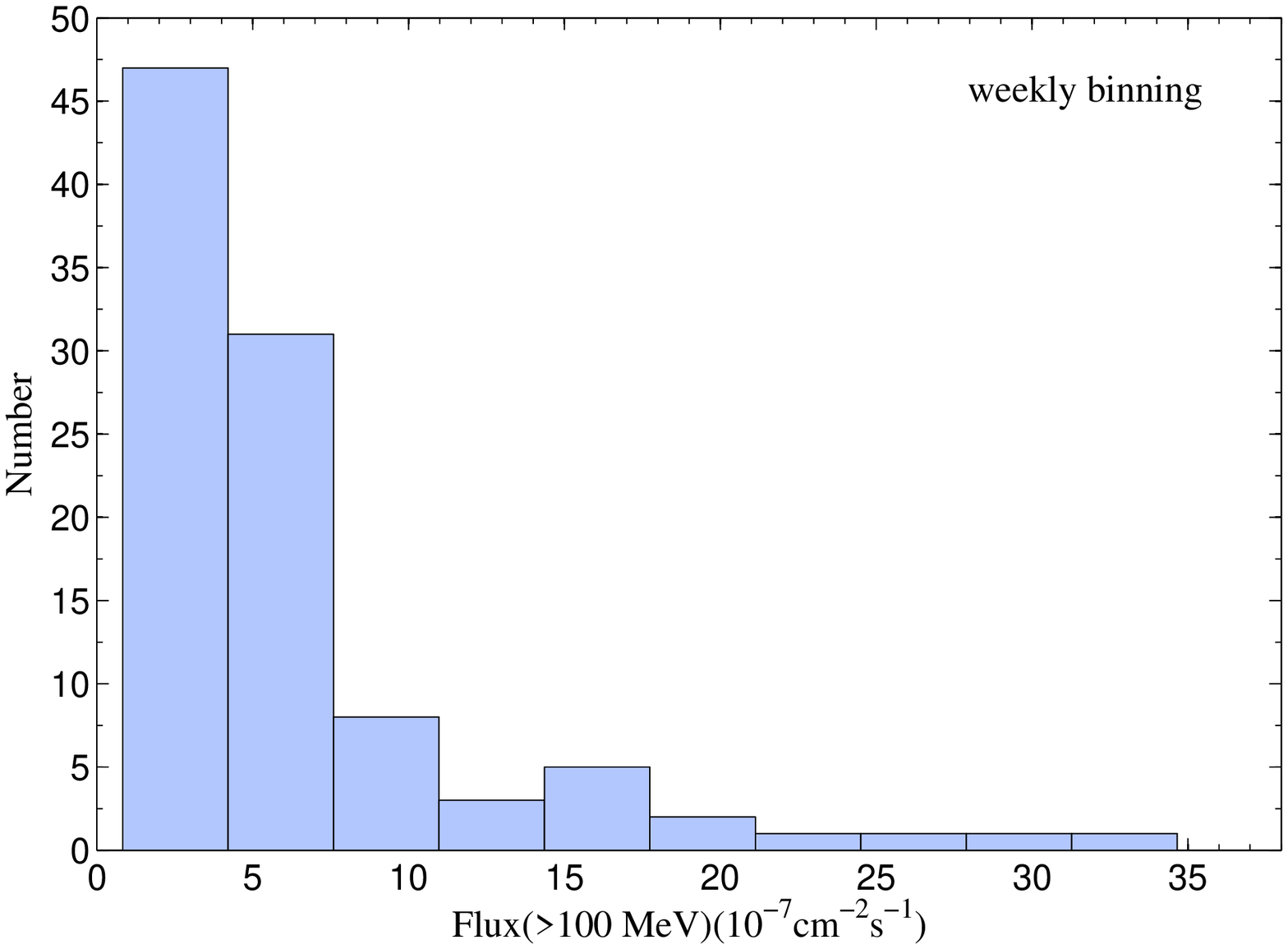}
\caption{Histograms of the synchrotron flux distribution
above 100 MeV, for monthly bin (left) and weekly bin
(right), respectively. In the left panel the Fermi/LAT
observation result is also shown (dashed histogram).
}
\label{fig:f4}
\end{figure}

To investigate fluctuations of emission in detail, we show in
Fig. \ref{fig:f3} the light curves of the simulated synchrotron
fluxes in the $\sim 100$ MeV energy band. Each bin of the light
curves represents an independent realization, and the length
of the time bin is proportional to the number of knots. We normalize
the time scale so that over a year there is an enhancement in flux by
at least a factor of 5-6, which may be responsible for a flare, and plot
the weekly and monthly bin lightcurves in the upper and lower panels of
Fig. \ref{fig:f3}, respectively. For comparison we also plot the Fermi/LAT
observed monthly light curve \citep{2010arXiv1011.3855F} in the
lower panel of Fig. \ref{fig:f3}. It can be seen that the
predicted fluctuation level of the simulated results is very
similar to that observed by Fermi/LAT. In order to show this more
clearly, we plot the histogram of the monthly bin flux
distribution for the simulated results along with that of Fermi/LAT
observation in the left panel of Fig. \ref{fig:f4}. These two
distributions are indeed similar. In the right panel of Fig.
\ref{fig:f4} the histogram of flux distribution for weekly bin is
also plotted. A concentration toward low-flux is evident, which is
distinct from a symmetric distribution around the average value.
This is because in the short time scale limit, the 
distribution of the gamma-ray fluxes is determined by the distributions 
of the knots (power-law) and the Doppler factors. Only the knots with 
both large enough sizes and high enough Doppler factors can produce 
large gamma-ray fluxes. Therefore the probability to give large fluxes 
is very small, and the flux distribution concentrates towards low-flux 
values. In the limit of long time scale, the distribution should 
be Gaussian as a result of central limit theorem. Such a prediction 
can be tested by a detailed analysis of the Fermi/LAT data.

\section{Conclusion and discussion}

In summary, we propose a statistical picture to explain the
observed $\gamma$-ray variability and flares of the Crab nebula,
using the fluctuations of the highest end of the electron spectra.
The electrons are thought to be accelerated in a series of knots,
with a size distribution $P(r_i)\propto r_i^{-\beta}$ and a
distribution of the Doppler factor. The maximal energy of the
electrons in the co-moving  frame is assumed to be proportional to
the size of the knots. Thus the rare knots with large sizes and
high Doppler boosts may generate electrons up to $\sim$PeV, and
hence, be responsible for the observed $\gamma$-rays flares at
Fermi/LAT. On the other hand, the low energy electrons are
generated by many smaller knots. The average effect smooths the
fluctuation, so that the low energy synchrotron component and the
IC component do not change significantly. This scenario can
naturally explain both the variability in the MeV-GeV band, and the
relatively steady emission in lower (optical, X-ray) and higher
(TeV) energies. The expected variability of the monthly bin
fluxes above 100 MeV are well consistent with that observed by
Fermi/LAT. The two large $\gamma$-ray flares can also be naturally
accounted for without additional assumptions.

In the simulation, we do not consider the detailed structure (e.g.
jet, torus) of the nebula. The knots are assumed to have a uniform
and isotropic distribution. Considering a more realistic
morphology of the nebula would not affect our conclusion
noticeably, as long as the distributions of the knot sizes and
Doppler factors are similar to those introduced here. In more general
terms, these knots may be interpreted as individual electron
acceleration events distributed throughout the nebula. We point out 
that at low energies contributions from large amount of knots are 
equivalent with previous studies that electrons are accelerated 
continuously inside the whole nebula. 
Although the model does not discuss the absolute time scales 
of flares and $\gamma$-ray fluctuations from the first principle, 
they may be related to timescales of developing MHD instabilities at
different spatial scales.

\acknowledgments We would like to thank Zhuo Li, Shuang-Nan Zhang,
Yi-Zhong Fan, Fang-Jun Lu and You-Jun Lu for helpful discussions. This 
work is partially supported by the Natural Sciences Foundation of China
grants 10773011, 11075169, 10633040 and 10921063, the 973 project grants
2010CB833000 and 2009CB824800, the NSF grant 
AST-0908362 and NASA grants NNX10AD48G and NNX10AP53G at UNLV.


\end{document}